\documentclass[letterpaper]{article}
\usepackage[utf8]{inputenc}

\usepackage{amsmath}
\usepackage{amsfonts}
\usepackage{amssymb}
\usepackage{amsthm}
\usepackage{authblk}
\usepackage{booktabs}
\usepackage{cite}
\usepackage{enumerate}
\usepackage{graphicx}
\usepackage[left=2cm,right=2cm,top=2cm,bottom=2cm]{geometry}
\usepackage{mathabx}
\usepackage{multirow}
\usepackage{subcaption}

\usepackage{color}

\newcommand{\es}[1]{\ensuremath{|(#1)^{(0)}\rangle}}

\author[1,2]{Pasquale Bosso\thanks{pasquale.bosso@uleth.ca}}
\author[1]{Saurya Das\thanks{saurya.das@uleth.ca}}
\affil[1]{Theoretical Physics Group and Quantum Alberta, University of Lethbridge,\protect\\ 4401 University Drive, Lethbridge, Alberta, Canada, T1K 3M4\vspace{1em}}
\affil[2]{Fakult\"at f\"ur Physik, Bielefeld University, D--33501 Bielefeld, Germany\vspace{1em}}

\title{Generalized ladder operators for the perturbed harmonic oscillator}

\date{}

\begin{document}

\maketitle

\begin{abstract}
In this paper, we construct corrections to the raising and lowering (i.e. ladder) operators for a quantum harmonic oscillator subjected to a polynomial type perturbation of any degree and to any order in perturbation theory.  
We apply our formalism to a couple of examples, namely $q$ and $p^4$ perturbations, and obtain the explicit form of those operators.
We also compute the expectation values of position and momentum for the above perturbations.
This construction is essential for defining coherent and squeezed states for the perturbed oscillator.
Furthermore, this is the first time that corrections to ladder operators for a harmonic oscillator with a generic perturbation and to an arbitrary order of perturbation theory have been constructed. 

\end{abstract}

\section{Introduction}

Perturbation methods in quantum mechanics have been used extensively in the area of Quantum Gravity Phenomenology.
It was in fact shown that the Generalized Uncertainty Principle (GUP) gives rise to small Planck-scale dependent terms in non-relativistic and relativistic Hamiltonians.
These terms can nevertheless have potential experimental implications \cite{Kempf1995_1,dasvagenas,Bosso2017a}. 
In particular, it was shown that there were $p^3$ and $p^4$ ($p=$ momentum, $q=$ position) Planck-scale corrections to the quantum Harmonic Oscillator (HO).
Such terms affect energy eigenvalues and eigenstates, that can be studied using standard perturbation techniques \cite{Bender1969}. 
In a recent paper \cite{Bosso2017a}, it was shown however that an alternative approach can be pursued.
Specifically, raising and lowering (or {\it ladder}) operators for the perturbed HO can be defined directly.
This not only offers a different perspective of the problem, but turns out to be essential in the definition of coherent and squeezed states for the perturbed oscillator.
It also proved advantageous to use these new operators in computing expectation values of various operators, which are quantities of interest for experiments.

In this paper we generalize the method developed in \cite{Bosso2017a}, and present a systematic procedure for defining ladder operators for an arbitrary selfadjoint perturbation in the form of a polynomial {\it of any degree} in $p$ and $q$, and to {\it any order} in perturbation theory.
Note that our treatment allows for terms with higher than second power in both position and momentum.
Therefore, it applies to a larger set of potentials than normally considered in the usual factorization method \cite{Infeld1951,Carinena,Ghosh,Carinena2017}.
As specific examples, we explicitly work out the case of perturbations of the form $q$ and $p^4$ and up to second order in perturbation theory.
Again, our results are essential to define coherent and squeezed states directly for a HO with generic perturbations.
It can also have potential applications in other fields, including in condensed matter physics and quantum field theory where perturbation theory plays a very important role.
As expected, our results reduce to the standard HO results when the perturbations are turned off.
This paper is organized as follows: in the next Section, we review some useful relations in perturbation theory.
In Section 3, we define the new set of ladder operators and give a prescription for evaluating them explicitly order by order.
We then use this to construct the operators  for the above examples in Section 4.
We conclude in Section 5. 

\section{Review of perturbation theory}

For the total Hamiltonian consisting of the unperturbed part $H_0$ and perturbation $\lambda V$, with perturbation parameter $\lambda$, i.e. \cite{Messiah}
\begin{equation}
	H = H_0 + \lambda V~. \label{eqn:pert_unpert}
\end{equation}
the eigenvalue and eigenvector equations are given order-by-order by ($i=1,2,\ldots$):
\begin{align}
	H_0 | E_i^{(0)} \rangle = & E_i^{(0)} | E_i^{(0)} \rangle, & H |E_i\rangle & = E_i |E_i\rangle~. \label{eqn:pert_eigen}
\end{align}
where
\begin{subequations}
\begin{align}
	E_i = & E_i^{(0)} + \lambda \epsilon_{i,(1)} + \lambda^2 \epsilon_{i,(2)} + \ldots = E_i^{(0)} + \sum_{n=1}^\infty \lambda^n \epsilon_{i,(n)}~, \\
	|E_i\rangle = & |E_i^{(0)}\rangle + \lambda |\eta_{i,(1)}\rangle + \lambda^2 |\eta_{i,(2)}\rangle + \ldots = |E_i^{(0)}\rangle + \sum_{n=1}^\infty \lambda^n |\eta_{i,(n)}\rangle~.
\end{align}
\end{subequations}
We follow the following convention here and in the rest of the paper:
$\epsilon_{i,(n)}$ and $|\eta_{i,(n)}\rangle$ are the $n$-th order corrections to the energy eigenvalue and eigenstate, respectively.
Therefore, a subscript in parenthesis indicates a correction term, while a superscript in parenthesis indicates a corrected quantity up to a given order.

Further, on imposing the following (standard) conditions \cite{Messiah}: 
\begin{equation}
	\langle E_i^{(0)} | E_i \rangle = \langle E_i^{(0)} | E_i^{(0)} \rangle = 1 \Rightarrow \langle E_i^{(0)} | \eta_{i,(n)} \rangle = 0 \qquad \forall n~, \label{eqn:orthogonality}
\end{equation}
one obtains the following results:
\begin{subequations}
\begin{align}
	\epsilon_{i,(n)} = & \langle E_i^{(0)} | V | \eta_{i,(n-1)} \rangle~, \label{eqn:nperturbation_value}\\
	|\eta_{i,(n)} \rangle = & \sum_{j \not= i} \frac{|E_j^{(0)} \rangle \langle E_j^{(0)} |}{E_i^{(0)} - E_j^{(0)}}[(V - \epsilon_{i,(1)}) | \eta_{i,(n-1)} \rangle - \epsilon_{i,(2)} | \eta_{i,(n-2)} \rangle - \ldots - \epsilon_{i,(n-1)} | \eta_{i,(1)} \rangle]~. \label{eqn:nperturbation_state}
\end{align}
\end{subequations}

For example, to first and second order in perturbation theory, one has respectively:
\begin{subequations}
\begin{align}
	\epsilon_{i,(1)} = & \langle E_i^{(0)} | V | E_i^{(0)} \rangle~,\\
	| \eta_{i,(1)} \rangle  = &
	\rlap{$\displaystyle{\sum_{j \not = i} \frac{ \langle E_j^{(0)} | V | E_i^{(0)} \rangle }{E_i^{(0)} - E_j^{(0)}} |E_j^{(0)} \rangle~,}$}
	\phantom{ \left[ \sum_{j \not = i} \sum_{k \not= i} \frac{ \langle E_j^{(0)} | V | E_k^{(0)} \rangle \langle E_k^{(0)} | V | E_i^{(0)} \rangle }{(E_i^{(0)} - E_j^{(0)}) (E_i^{(0)} - E_k^{(0)})} - \sum_{j \not = i}\frac{ \langle E_j^{(0)} | V | E_i^{(0)} \rangle \langle E_i^{(0)} | V | E_i^{(0)} \rangle }{(E_i^{(0)} - E_j^{(0)})^2} \right] |E_j^{(0)} \rangle~.} \label{eqn:1perturbation}
\end{align}
\end{subequations}
\begin{subequations}
\begin{align}
	\epsilon_{i,(2)} = & \sum_{j \not = i} \frac{|\langle E_j^{(0)} | V | E_i^{(0)} \rangle |^2 }{E_i^{(0)} - E_j^{(0)}}~,\\
	| \eta_{i,(2)} \rangle = & \left[ \sum_{j \not = i} \sum_{k \not= i} \frac{ \langle E_j^{(0)} | V | E_k^{(0)} \rangle \langle E_k^{(0)} | V | E_i^{(0)} \rangle }{(E_i^{(0)} - E_j^{(0)}) (E_i^{(0)} - E_k^{(0)})} - \sum_{j \not = i}\frac{ \langle E_j^{(0)} | V | E_i^{(0)} \rangle \langle E_i^{(0)} | V | E_i^{(0)} \rangle }{(E_i^{(0)} - E_j^{(0)})^2} \right] |E_j^{(0)} \rangle~.
\end{align}
\end{subequations}

One can also show that the eigenstate corrected to the second order is not normalized.
To see this, we first notice that \eqref{eqn:orthogonality} implies $\langle \eta_{i,(1)} | E_i^{(0)} \rangle = \langle \eta_{i,(2)} | E_i^{(0)} \rangle = 0$.
Furthermore, $\langle \eta_{i,(1)} | \eta_{i,(2)} \rangle$ and $\langle \eta_{i,(2)} | \eta_{i,(2)} \rangle$ are terms of orders higher than second, $\lambda^3$ and $\lambda^4$, respectively.
Therefore the only non-zero terms are $\langle E_i^{(0)} | E_i^{(0)} \rangle = 1$ and $\lambda^2 \langle \eta_{i,(1)} | \eta_{i,(1)} \rangle$.
In particular, this last term is of order $\lambda^2$ and needs to be retained in the second order expansion.
In general, a perturbed state up to order $m\geq 2$ is not normalized.

\section{Ladder operators for the perturbed harmonic oscillator}

In this Section, we first show that a unique set of ladder operators exists for a generic perturbation of the HO.
We then give an explicit construction of the operators to any order in perturbation theory.
We start by assuming that there exists a Hermitian operator $C$, with complete set of eigenvectors $|\gamma\rangle$ and corresponding (real) eigenvalues $\gamma$.
We assume that the eigenvalues are non-degenerate, without loss of generality.  
Then the set $\{|\gamma\rangle \}$ forms an orthonormal basis for the Hilbert space $\mathcal{H}$, \emph{i.e.}, 
\begin{equation}
	C |\gamma\rangle = \gamma |\gamma\rangle \qquad \mbox{for} \qquad |\gamma\rangle \in \mathcal{H}~, \quad \mbox{and} \quad \langle \delta | \gamma \rangle = \left\{ \begin{array}{cc}
  	1 & \mbox{for } \delta=\gamma~,\\
    0 & \mbox{for } \delta\not=\gamma~,
  \end{array}\right. \quad \gamma, \delta \in \mathbb{R}~. \label{def:C}
\end{equation}
Let us, then, define a new operator, $A$, such that $A |\gamma\rangle = \sqrt{\gamma} |\gamma - 1\rangle$.
It is easy to prove that the following relations hold
\begin{align}
		A^\dagger |\gamma\rangle & = \sqrt{\gamma + 1} |\gamma + 1\rangle~, & C = & A^\dagger A~, & [A,A^\dagger] & = 1~. \label{eqn:prop1}
\end{align}
Notice that
the conditions $C=A^\dagger A$ and the fact that the vectors in the Hilbert space have non-negative norms require $\gamma\in\mathbb{N}$.

Next, we consider a perturbed HO Hamiltonian, where the perturbation is a selfadjoint polynomial in $q$ and $p$ of arbitrary degree
\begin{equation}
	H = H^{(0)} + \lambda V(q,p)~.
\end{equation}
We then define two operators, $\tilde{a}$ and $\tilde{N}$, that act like an annihilation and a number operators for the perturbed eigenstates
\begin{subequations}
\begin{align}
	\tilde{a} |n\rangle = & \sqrt{n} |n-1\rangle~, & H \tilde{a} |n\rangle = & E_{n-1} \tilde{a} |n\rangle~, \label{def:tilde_a}\\
	\tilde{N} |n\rangle = & n |n\rangle~, & H \tilde{N} |n\rangle = & E_n \tilde{N} |n\rangle~,
\end{align}
\end{subequations}
with $H |n\rangle = E_n |n\rangle$ and where $E_{n}$ is the eigenvalue of the perturbed Hamiltonian $H$ for the perturbed eigenstate $|n\rangle$.
Furthermore, from \eqref{eqn:prop1} with $A\equiv \tilde{a}$ and $C \equiv \tilde{N}$, we have the following additional relations
\begin{align}
	\tilde{a}^\dagger |n\rangle = & \sqrt{n+1} |n+1\rangle~, & H \tilde{a}^\dagger |n\rangle = & E_{n+1} \tilde{a}^\dagger |n\rangle~,\label{def:tilde_a_dagger}\\
	\tilde{N} = & \tilde{a}^\dagger \tilde{a}~,\\
	[\tilde{a},\tilde{a}^\dagger] = & 1~.
\end{align}
As required for the perturbed eigenstates and eigenvalues, we also require 
\begin{equation}
	\lim_{\lambda \rightarrow 0} \tilde{a} = a~,
\end{equation}
that is, we recover the standard annihilation operator for the HO once the perturbation is turned off.
If we assume that the perturbation $\lambda V$ is small, we can expand $\tilde{a}$ in series
\begin{equation}
	\tilde{a} = \tilde{a}^{(0)} + \lambda \alpha_{(1)} + \ldots = \sum_{m=0}^\infty \lambda^m \alpha_{(m)}~,
\end{equation}
where by definition $\tilde{a}^{(0)} \equiv \alpha_{(0)} \equiv a$ and where $\alpha_{(n)}$ is the $n$-th order correction of the annihilation operator.
It is important to notice that,
\begin{enumerate}[(i)]
\item Following our convention, $\tilde{a}^{(m)}$ is the corrected annihilation operator up to order $m$, and 
\item in the same way that $|n\rangle$ is a normalized state but $|n^{(m)}\rangle$ is not in general, $\tilde{a}$ is the correct annihilation operator for normalized states $|n\rangle$ while $\tilde{a}^{(m)}$ is the correct annihilation operator for the (in general) non-normalized states $|n^{(m)}\rangle$.
\end{enumerate}

	Now, using the definition \eqref{def:tilde_a}, we find
	\begin{equation}
		\tilde{a} |n\rangle = \left[ a + \sum_{m=1}^\infty \lambda^m \alpha_{(m)}\right] \left[|n^{(0)}\rangle + \sum_{m=1}^\infty \lambda^m |\eta_{n,(m)}\rangle\right] ~.
	\end{equation}
	Analyzing this equation order by order, we find
	\begin{subequations}
	\begin{align}
		\sqrt{n} |\eta_{n-1,(1)}\rangle = & \left[a |\eta_{n,(1)}\rangle + \alpha_{(1)} |n^{(0)}\rangle\right] ~, \tag{$\theequation^{(1)}$} \\
		\sqrt{n} |\eta_{n-1,(2)}\rangle = & \left[a |\eta_{n,(2)}\rangle + \alpha_{(1)} |\eta_{n,(1)}\rangle + \alpha_{(2)} |n^{(0)}\rangle\right] ~, \tag{$\theequation^{(2)}$} \\
		\vdots & \nonumber\\
		\sqrt{n} |\eta_{n-1,(m)}\rangle = & \left[a |\eta_{n,(m)}\rangle + \alpha_{(1)} |\eta_{n,(m-1)}\rangle + \ldots + \alpha_{(m)} |n^{(0)}\rangle\right]~, \tag{$\theequation^{(m)}$}\\
		\vdots & \nonumber
	\end{align}
	\end{subequations}
	In general, we have the following iterative relation for the corrections to $\tilde{a}$ to any order in perturbation theory
	\begin{equation}
		\alpha_{(m)} |n^{(0)}\rangle = \sqrt{n} |\eta_{n-1,(m)}\rangle - a |\eta_{n,(m)}\rangle - \alpha_{(1)} |\eta_{n,(m-1)}\rangle - \ldots - \alpha_{(m-1)} |\eta_{n,(1)}\rangle~. \label{eqn:correction_annihilation}
	\end{equation}
Note that the corrections are unique to any order in perturbation theory and that it does not depend on the particular level it applies to.

\section{Explicit results for perturbations to first and second order}

Following \eqref{eqn:correction_annihilation}, 
we can now find explicit expressions for the first and second order corrections 
to the annihilation, creation, and number operators.

\subsection{First order correction}

From \eqref{eqn:correction_annihilation}, the first order correction to the annihilation operator is given by
\begin{equation}
	\alpha_{(1)} |n^{(0)}\rangle = \sqrt{n} |\eta_{n-1,(1)}\rangle - a |\eta_{n,(1)}\rangle~.
\end{equation}
Using \eqref{eqn:1perturbation}, we find
\begin{equation}
	\hbar \omega \alpha_{(1)} |n^{(0)}\rangle 
	= \sum_{j \not = n} \frac{ \langle (j-1)^{(0)} | [V, a] | n^{(0)} \rangle }{n - j} |(j-1)^{(0)} \rangle ~.
\end{equation}
The operator $V$ contains a sum of products of $a$ and $a^\dagger$, not necessarily ordered.
Let us consider one of such terms, containing $b$ copies of $a^\dagger$ and $c$ copies of $a$.
We represent this term by the symbol $V_c^b$.
It is then easy to see that for each of these terms we find
	\begin{equation}
		\sum_m |m^{(0)}\rangle \langle m^{(0)}| V_c^b | n^{(0)} \rangle = V_c^b | n^{(0)} \rangle~.
	\end{equation}
Using this property, we have
\begin{equation}
	\sum_{j \not = n} \frac{ \langle (j-1)^{(0)} | [V_c^b,a] | n^{(0)} \rangle }{n - j} |(j-1)^{(0)} \rangle = \frac{[V_c^b,a] |n^{(0)}\rangle}{c-b} = [\overline{V}_c^b,a] |n^{(0)}\rangle , \qquad \mbox{with} \qquad c-b \not = 0~, \label{eqn:gen_term_1ord}
\end{equation}
where
\begin{equation}
	\left\{ \begin{array}{cc}
		\overline{V}_c^b = \displaystyle{\frac{V_c^b}{c-b}}~, & \mbox{for } \qquad b\not=c\\
		[1em]\overline{V}_c^b = 0~, & \mbox{for } \qquad b = c
	\end{array} \right.
\end{equation}
The terms with $b=c$ can be safely excluded from the definition of $\overline{V}_c^b$,
since these terms do not have any role in the above expressions.
Such terms are indeed automatically excluded in the sum \eqref{eqn:gen_term_1ord}.
In this way, we find
\begin{equation}
	\alpha_{(1)} = \frac{1}{\hbar \omega} [\overline{V}, a]	~,
\end{equation}
where $\overline{V}$ is the sum of all the terms $\overline{V}_c^b$.
 
Notice that, since $V$ is a Hermitian operator, 
the operator $\overline{V}$ is anti-Herminian, \emph{i.e.}, $\overline{V}^\dagger = - \overline{V}$.
Following similar arguments as above, it is easy to prove that
	\begin{equation}
		\alpha_{(1)}^\dagger = \frac{1}{\hbar \omega} [\overline{V},a^\dagger]~. \label{eqn:1ord_creation}
	\end{equation}
As for the number operator, we can define the following perturbation series
\begin{equation}
	\tilde{N} = \sum_{m=0}^\infty \lambda^m \nu_{(m)}~, \qquad \mbox{with} \qquad \nu_{(0)} = \tilde{N}^{(0)} = N~,
\end{equation}
finding
	\begin{equation}
		\nu_{(1)} = a^\dagger \alpha_{(1)} + \alpha_{(1)}^\dagger a = \frac{1}{\hbar \omega} [\overline{V}, N] ~. \label{eqn:1ord_number}
	\end{equation}

\subsubsection{Example: $V=q$} \label{sssec:ex_q1}

In this case the perturbation is given by $V = q = \sqrt{\hbar/2 m \omega} (a + a^\dagger)$.
We have the following terms
\begin{align}
	V_1^0 = & a~, & V_0^1 = & a^\dagger~, & \mbox{and}&& \overline{V}_1^0 = & a~, & \overline{V}_0^1 = & - a^\dagger~.
\end{align}
Therefore, we then define
\begin{equation}
	\overline{V} = \sqrt{\frac{\hbar}{2 m \omega}} (a - a^\dagger)~, \qquad \mbox{from which} \qquad [\overline{V},a] = \sqrt{\frac{\hbar}{2 m \omega}}~.
\end{equation}
Therefore
\begin{equation}
	\alpha_{(1)} = \frac{1}{\sqrt{2 \hbar m \omega^3}}~, \qquad \mbox{and} \qquad \tilde{a}^{(1)} = a + \frac{1}{\sqrt{2 \hbar m \omega^3}}~. \label{eqn:1order_a}
\end{equation}
The corresponding corrections to the creation and number operators follow from \eqref{eqn:1ord_creation} and \eqref{eqn:1ord_number}.
We also have that
\begin{equation}
	|\eta_{n,(1)}\rangle = \sum_{j \not = n} \frac{ \langle j^{(0)} | V | n^{(0)} \rangle }{\hbar \omega (n - j)} |j^{(0)} \rangle = \frac{1}{\sqrt{2 \hbar m \omega^3}} \left( \sqrt{n} |(n-1)^{(0)}\rangle - \sqrt{n+1} |(n+1)^{(0)}\rangle \right)~.
\end{equation}

We can easily invert \eqref{eqn:1order_a}, finding
\begin{align}
	a = & \tilde{a}^{(1)} - \frac{\lambda}{\sqrt{2 \hbar m \omega^3}}~, & a^\dagger = & \tilde{a}^{(1)} {}^\dagger - \frac{\lambda}{\sqrt{2 \hbar m \omega^3}}~.
\end{align}
We can use these relations to find, for example, the form of the position operator in terms of the new operators
\begin{align}
	q = & \tilde{q}^{(1)} - \frac{\lambda}{m \omega^2}~, & & \mbox{where} & \tilde{q}^{(m)} = & \sqrt{\frac{\hbar}{2 m \omega}} (\tilde{a}^{(m)} + \tilde{a}^{(m)} {}^\dagger)~.
\end{align}
Then the mean value of the position on an energy eigenstate, up to first order, is
\begin{equation}
	\langle n^{(1)} | q | n^{(1)} \rangle = - \frac{\lambda}{m \omega^2}~.
\end{equation}
Similarly for the momentum operator, we find
\begin{align}
	p = & \tilde{p}^{(1)}~, & & \mbox{where} & \tilde{p}^{(m)} = & i \sqrt{\frac{\hbar m \omega}{2}} (\tilde{a}^{(m) \dagger} - \tilde{a}^{(m)})~,
\end{align}
therefore
\begin{equation}
	\langle n^{(1)} | p | n^{(1)} \rangle = 0~.
\end{equation}

\subsubsection{Example: $V=p^4$} \label{sssec:ex_p4}

In this case the perturbation is
\begin{equation}
	V = p^4 = \left(\frac{\hbar m \omega}{2}\right)^2 \left[a^4 - 2 a (2 N + 1) a + 3 (2 N^2 + 2 N + 1 ) - 2 a^\dagger (2 N + 1) a^\dagger + a^\dagger {}^4\right]~,
\end{equation}
from which we have
\begin{equation}
	\overline{V} = \left(\frac{\hbar m \omega}{2}\right)^2 \left[\frac{a^4}{4} - a (2 N + 1) a + a^\dagger (2 N + 1) a^\dagger - \frac{a^\dagger {}^4}{4}\right]~, 
	\qquad [\overline{V},a] = \left(\frac{\hbar m \omega}{2}\right)^2 \left(2 a^3 - 6 N a^\dagger + a^\dagger {}^3\right)~,
\end{equation}
therefore
\begin{equation}
	\tilde{a}^{(1)} = a + \lambda \frac{\hbar m^2 \omega}{4} \left(2 a^3 - 6 N a^\dagger + a^\dagger {}^3\right)~. \label{eqn:1order_a_q4}
\end{equation}
As for the correction to the energy eigenstate, we find
\begin{multline}
	|\eta_{n,(1)}\rangle = \frac{\hbar m^2 \omega}{4} \left[\frac{\sqrt{n^{\underline{4}}}}{4} \es{n-4}
		- \sqrt{n^{\underline{2}}} (2 n - 1) \es{n-2}
		+ \sqrt{(n+1)^{\overline{2}}} (2 n + 3) \es{n+2} \right. \\
	\left. - \frac{\sqrt{(n+1)^{\overline{4}}}}{4} \es{n+4}\right]~.
\end{multline}

Inverting \eqref{eqn:1order_a_q4} we find
\begin{align}
	a = & \tilde{a}^{(1)} - \lambda \frac{\hbar m^2 \omega}{4} \left(2 a^{(1)} {}^3 - 6 N^{(1)} a^{(1)} {}^\dagger + a^{(1)} {}^{\dagger 3}\right)~, & 
	a^\dagger = & \tilde{a}^{(1)} {}^\dagger - \lambda \frac{\hbar m^2 \omega}{4} \left(2 a^{(1)} {}^{\dagger 3} - 6 a^{(1)} {} N^{(1)} + a^{(1)} {}^3\right)
\end{align}
where we used $\lambda a = \lambda \tilde{a}^{(1)}$, up to first order in perturbation theory.
For the position operator we then find
\begin{equation}
	q = \tilde{q}^{(1)} - \frac{3}{4} \lambda \sqrt{\frac{\hbar^3 m^3 \omega}{2}} \left[ a^{(1)} {}^3 - 2 ( a^{(1)} {} N^{(1)} + N^{(1)} a^{(1)} {}^\dagger ) + a^{(1)} {}^{\dagger 3} \right]~.
\end{equation}
For its mean value we have
\begin{equation}
	\langle n^{(1)} | q | n^{(1)} \rangle = 0~.
\end{equation}
Furthermore, we have
\begin{align}
	p = & p^{(1)} - i \frac{\lambda}{4} \sqrt{\frac{\hbar^3 m^5 \omega^3}{2}} \left[a^{(1)} {}^{\dagger 3} - 6 ( a^{(1)} {} N^{(1)} + N^{(1)} a^{(1)} {}^\dagger ) - a^{(1)} {}^3 \right]~, & & \Rightarrow & \langle n^{(1)} | p | n^{(1)} \rangle = & 0~.
\end{align}

\subsection{Second order correction}

Again from \eqref{eqn:correction_annihilation}, we have
\begin{multline}
	(\hbar \omega)^2 \alpha_{(2)} |n^{(0)}\rangle
	= \sum_{j \not = n} \sum_{k \not= n} \frac{ |(j-1)^{(0)} \rangle \langle (j-1)^{(0)} | \left[V | (k-1)^{(0)} \rangle \langle (k-1)^{(0)} | V a - a V | k^{(0)} \rangle \langle k^{(0)} | V \right] | n^{(0)} \rangle }{(n - j) (n - k)} \\
	- \sum_{j \not = n}\frac{ |(j-1)^{(0)} \rangle \langle (j-1)^{(0)} | \left[ \langle (n-1)^{(0)} | V | (n-1)^{(0)} \rangle V a - \langle n^{(0)} | V | n^{(0)} \rangle a V \right] | n^{(0)} \rangle }{(n - j)^2} \\
	- [\overline{V},a] \sum_{j \not = n} \frac{ \langle j^{(0)} | V | n^{(0)} \rangle }{n - j} |j^{(0)} \rangle ~.
\end{multline}
It is convenient to introduce a new operator, $\widecheck{V}$, containing only terms of $V$ with equal number of annihilation and creation operators.
In other words, if $V_c^b$ is a generic term in $V$, the corresponding term in $\widecheck{V}$ is
\begin{equation}
	\widecheck{V}_c^b = \left\{\begin{array}{cc}
		V_c^b & \mbox{for } b=c~,\\
		0 & \mbox{for } b\not=c~,
 	\end{array}\right.
\end{equation}
where we used the same notation of the previous subsection.
With this definition, we have
	\begin{align}
		a |n^{(0)}\rangle \langle n^{(0)}|V|n^{(0)}\rangle & = a \widecheck{V} |n^{(0)}\rangle~, &
		a |n^{(0)}\rangle \langle (n-1)^{(0)}|V|(n-1)^{(0)}\rangle & = \widecheck{V} a |n^{(0)}\rangle~.
	\end{align}
Similarly as before, we consider two of the terms composing $V$, namely $V_c^b$ and $V_{c'}^{b'}$.
For this case, we find
\begin{equation}
	\alpha_{(2)} = \frac{1}{(\hbar \omega)^2} \left\{\left[ \overline{V \overline{V}}, a \right]
	- \left[ \overline{\overline{V}} \widecheck{V}, a \right]
	- [\overline{V},a] \overline{V} \right\}~. \label{eqn:correction2}
\end{equation}

As done for the first order correction, we can find expressions for the second order corrections of the creation and number operators.
Unlike the first order, in this case it will not be possible to find a common expression for all the operators.
The reason is in the fact that, although many of the terms in \eqref{eqn:correction2} have defined hermiticity properties, the term $\overline{V \overline{V}}$ does not.
In fact, we see that $\overline{\overline{V}}$ and $\widecheck{V}$ are Hermitian operators when $V$ is Hermitian.
More in general, the operation $\widecheck{\cdot}$ will always generate a Hermitian operator, while the multiple applications of the operation $\overline{\cdot}$ on an Hermitian operator produces alternatively anti-Hermitian and Hermitian operators.
With these properties, we find
\begin{equation}
	\alpha_{(2)}^\dagger = \frac{1}{(\hbar \omega)^2} \left\{- \left[ \left(\overline{V \overline{V}}\right)^\dagger, a^\dagger \right]
	+ \left[ \widecheck{V} \overline{\overline{V}}, a^\dagger \right]
	+ \overline{V} [\overline{V},a^\dagger] \right\}~,
\end{equation}
and
\begin{equation}
	\nu_{(2)} = \alpha^\dagger_{(1)} \alpha_{(1)} + a^\dagger \alpha_{(2)} + \alpha_{(2)}^\dagger a = \frac{1}{(\hbar \omega)^2} \left\{\frac{1}{2} \overline{V} [\overline{V}, N]
	- a^\dagger \left( \frac{1}{2} [\overline{V}^2, a] - [\overline{V \overline{V}} , a] + \left[\overline{\overline{V}} \widecheck{V} , a\right] \right) \right\} + \mbox{h.c.}
\end{equation}

\subsubsection{Example: $V=q$}

Consider the same case of Example \ref{sssec:ex_q1}.
For this perturbation we have
\begin{subequations}
\begin{align}
	\overline{V} = & \sqrt{\frac{\hbar}{2 m \omega}} (a - a^\dagger)~, & 
	\overline{V \overline{V}} = & \frac{\hbar}{2 m \omega} \left(\frac{a^2}{2} + \frac{a^\dagger {}^2}{2}\right)~, & 
	\overline{\overline{V}} = & \sqrt{\frac{\hbar}{2 m \omega}} (a + a^\dagger)~, & 
	\widecheck{V} = & 0~,\\
	\left[ \overline{V \overline{V}}, a \right] = & - \frac{\hbar}{2 m \omega} a^\dagger~, &
	\left[ \overline{\overline{V}} \widecheck{V}, a \right] = & 0~, & 
	[\overline{V},a] \overline{V} = & \frac{\hbar}{2 m \omega} (a - a^\dagger)~,
\end{align}
\end{subequations}
obtaining
\begin{equation}
	\alpha_{(2)} = - \frac{1}{2 \hbar m \omega^3} a~.
\end{equation}
Therefore, for the annihilation operator up to second order in perturbation theory, we have
\begin{equation}
	\tilde{a}^{(2)} = a + \lambda \frac{1}{\sqrt{2 \hbar m \omega^3}} - \lambda^2 \frac{1}{2 \hbar m \omega^3} a~. \label{eqn:2order_a_q}
\end{equation}
As for the second order correction to the eigenstates, we find
\begin{multline}
	|\eta_{n,(2)}\rangle = \frac{1}{2 \hbar m \omega^3} \sum_{j \not = n} \sum_{k \not= n} \frac{ \langle j^{(0)} | (a + a^\dagger) | k^{(0)} \rangle \langle k^{(0)} | (a + a^\dagger) | n^{(0)} \rangle }{(n - j) (n - k)} |j^{(0)} \rangle = \\
	= \frac{1}{2 \hbar m \omega^3} \sum_{j \not = n} \left[\sqrt{n} \frac{ \langle j^{(0)} | (a + a^\dagger) | (n-1)^{(0)} \rangle }{(n - j) } |j^{(0)} \rangle - \sqrt{n+1} \frac{ \langle j^{(0)} | (a + a^\dagger) | (n+1)^{(0)} \rangle }{(n - j) } |j^{(0)} \rangle\right]= \\
	= \frac{1}{2 \hbar m \omega^3} \left[\frac{\sqrt{n^{\underline{2}}}}{2} |(n-2)^{(0)} \rangle + \frac{\sqrt{(n+1)^{\overline{2}}}}{2} |(n+2)^{(0)} \rangle\right]~.
\end{multline}
Hence, the perturbed energy eigenstate, up to second order in perturbation theory, is
\begin{multline}
	|n^{(2)}\rangle = |n^{(0)}\rangle + \lambda \frac{1}{\sqrt{2 \hbar m \omega^3}} \left[ \sqrt{n} |(n-1)^{(0)}\rangle - \sqrt{n+1} |(n+1)^{(0)}\rangle \right] \\
	+ \lambda^2 \frac{1}{2 \hbar m \omega^3} \left[\frac{\sqrt{n^{\underline{2}}}}{2} |(n-2)^{(0)} \rangle + \frac{\sqrt{(n+1)^{\overline{2}}}}{2} |(n+2)^{(0)} \rangle\right]~.
\end{multline}

Inverting \eqref{eqn:2order_a_q} we find
\begin{align}
	a = & \tilde{a}^{(2)} - \lambda \frac{1}{\sqrt{2 \hbar m \omega^3}} + \lambda^2 \frac{1}{2 \hbar m \omega^3} \tilde{a}^{(2)}~, 
    & a^\dagger = & \tilde{a}^\dagger {}^{(2)} - \lambda \frac{1}{\sqrt{2 \hbar m \omega^3}} + \lambda^2 \frac{1}{2 \hbar m \omega^3} \tilde{a}^\dagger {}^{(2)}~,
\end{align}
where we noticed that $\lambda^2 a = \lambda^2 \tilde{a}^{(2)}$ up to second order in $\lambda$.
As for the position operator, we find
\begin{equation}
	q = \tilde{q}^{(2)} - \frac{\lambda}{2 m \omega^2} + \frac{\lambda^2}{\sqrt{2^3 \hbar m^3 \omega^7}} \tilde{q}^{(2)}~.
\end{equation}
Since the norm of the perturbed eigenstate up to second order is
\begin{equation}
	\langle n^{(2)} | n^{(2)} \rangle = 1 + \frac{\lambda^2}{2 \hbar m \omega^3}(2n + 1)~,
\end{equation}
we have for the average position
\begin{equation}
	\frac{\langle n^{(2)} | q | n^{(2)} \rangle}{\langle n^{(2)} | n^{(2)} \rangle} \simeq - \frac{\lambda}{2 m \omega^2}~.
\end{equation}
As for the momentum, we find
\begin{align}
	p = & \tilde{p}^{(2)} + i \lambda^2 \sqrt{\frac{\hbar}{2^3 m \omega^5}} \tilde{p}^{(2)} & & \Rightarrow & \frac{\langle n^{(2)} | p | n^{(2)} \rangle}{\langle n^{(2)} | n^{(2)} \rangle} = & 0~.
\end{align}

\subsubsection{Example: $V=p^4$}

Consider the case of Example \ref{sssec:ex_p4}.
We have
\begin{subequations}
\begin{align}
	\overline{V} = & \left(\frac{\hbar m \omega}{2}\right)^2 \left[ \frac{a^4}{4} - a (2N + 1) a + 2 a^\dagger (2N + 1) a^\dagger - \frac{a^\dagger {}^4}{4} \right]~,\\
	\overline{V \overline{V}} = & \left(\frac{\hbar m \omega}{2}\right)^4 \frac{1}{2} \left[ \frac{a^8}{16}
	- a^3 \left(N + \frac{7}{6}\right) a^3
	+ \frac{1}{4} a^2 \left( 19 N^2 + 7 N - \frac{9}{2}\right) a^2
	- a \left( 11 N^3 - \frac{51}{2} N^2 - \frac{61}{2} N - 30 \right) a \right. \nonumber \\
	& \left. - a^\dagger \left( 11 N^3 + \frac{117}{2} N^2 + \frac{107}{2} N + 36 \right) a^\dagger
	+ \frac{1}{4} a^\dagger {}^2 \left( 19 N^2 + 31 N + \frac{15}{2} \right) a^\dagger {}^2
	- a^\dagger {}^3 \left( N - \frac{1}{6} \right) a^\dagger {}^3
	+ \frac{a^\dagger {}^8}{16} \right] ~,\\
	\overline{\overline{V}} = & \left(\frac{\hbar m \omega}{2}\right)^2 \frac{1}{2} \left[\frac{a^4}{8}
	- a \left( 2 N + 1 \right) a
	- a^\dagger \left( 2 N + 1 \right) a^\dagger
	+ \frac{a^\dagger {}^4}{8}\right]~,\\
	\widecheck{V} = & \left(\frac{\hbar m \omega}{2}\right)^2 3 (2 N^2 + 2 N + 1 )~,\\
	\overline{\overline{V}}\widecheck{V} = & \left(\frac{\hbar m \omega}{2}\right)^4 3 \left[\frac{1}{8} a^2 \left( N^2 + 5 N + \frac{13}{2}\right) a^2
	- a \left( 2 N^3 + 7 N^2 + 8 N + \frac{5}{2} \right) a
	- a^\dagger \left( 2 N^3 - N^2 + \frac{1}{2} \right) a^\dagger \right. \nonumber \\
	& \left. + \frac{1}{8} a^\dagger {}^2 \left(N^2 - 3 N + \frac{5}{2} \right) \right]~, \\
	[\overline{V \overline{V}},a] = & \left(\frac{\hbar m \omega}{2}\right)^4 \left[ \frac{a^7}{2} 
	- \frac{1}{2} a^3 \left( \frac{19}{2} N - 3\right) a^2 
	+ 3 a^2 \left( \frac{11}{2} N^2 - 14 N + 1 \right) a
	+ \left(\frac{55}{2} N^3 + 84 N^2 + \frac{29}{2} N + 33\right) a^\dagger \right. \nonumber \\
	& \left. - \frac{3}{2} a^\dagger \left( \frac{19}{2} N^2 + 5 N + \frac{1}{2} \right) a^\dagger {}^2
	+ a^\dagger {}^2 \left(\frac{7}{2} N - 2 \right) a^\dagger {}^3
	- \frac{a^\dagger {}^7}{4} \right] ~,\\
	[\overline{\overline{V}} \widecheck{V},a] = & \left(\frac{\hbar m \omega}{2}\right)^4 3 \left[ - \frac{1}{2} a^3 \left( \frac{N}{2} + 1 \right) a^2
	+ a^2 \left( 6 N^2 + 8 N + 3 \right) a
	+ \left(10 N^3 - 16 N^2 + 11 N - 2 \right) a^\dagger \right. \nonumber \\
	& \left. - \frac{1}{2} a^\dagger \left( \frac{3}{2} N^2 - 5 N + \frac{9}{2} \right) a^\dagger {}^2 \right] ~,\\
	[\overline{V},a] \overline{V} = & \left(\frac{\hbar m \omega}{2}\right)^4 \left[ \frac{a^7}{2}
	- 2 a^3 \left( 2 N + 3 \right) a^2
	- 3 a^2 \left( \frac{N^2}{2} - 2 N + 2 \right) a
	+ a \left( \frac{65}{4} N^3 - \frac{27}{2} N^2 + \frac{211}{4} N + \frac{9}{2}\right) \right. \nonumber \\
	& \left. - \frac{a^\dagger {}^7}{4}
	+ \frac{7}{2} a^\dagger {}^2 N a^\dagger {}^3
	- 6 a^\dagger \left( 2 N^2 + N - 1 \right) a^\dagger {}^2
	- \left( \frac{5}{2} N^3 - 6 N^2 + \frac{37}{2} N - 3\right) a^\dagger \right] ~.
\end{align}
\end{subequations}
For the second order correction to the annihilation operator we then have
\begin{multline}
	\alpha_{(2)} = \left(\frac{\hbar^2 m^4 \omega^2}{16}\right) \left[9 a^5
	- 72 a^2 N a
	- \frac{1}{2} a \left( \frac{65}{2} N^3 - 27 N^2 + \frac{211}{2} N + 9 \right)
	+ 18 \left( 7 N^2 + 2 \right) a^\dagger
	- 9 a^\dagger N a^\dagger {}^2
	- 2 a^\dagger {}^5 \right]~.
\end{multline}

As for the correction to the eigenstate, we find
\begin{multline}
	|\eta_{n,(2)}\rangle =  \frac{\hbar^2 m^4 \omega^2}{16} \left[\frac{\sqrt{n^{\underline{8}}}}{32} \es{n-8}
		+ \frac{\sqrt{n^{\underline{6}}}}{2} \left( n - \frac{11}{6} \right) \es{n-6}
		+ \sqrt{n^{\underline{4}}} \left(2 n^2 - 9 n + 7\right) \es{n-4} \right. \\
		- \frac{\sqrt{n^{\underline{2}}}}{4} \left(2 n^3 + 129 n^2 - 107 n + 66\right) \es{n-2}
		- \frac{\sqrt{(n+2)^{\overline{2}}}}{4} \left(2 n^3 - 123 n^2 - 359 n - 300\right) \es{n+2} \\
		\left. + \sqrt{(n+1)^{\overline{4}}} \left(2 n^2 + 13 n + 18\right) \es{n+4}
		+ \frac{\sqrt{(n+1)^{\overline{6}}}}{2} \left(n + \frac{17}{6}\right) \es{n+6}
		+ \frac{\sqrt{n^{\overline{8}}}}{32} \es{n+8} \right]~.
\end{multline}

Writing $a$ and $a^\dagger$ in terms of $\tilde{a}^{(2)}$ and $\tilde{a}^{(2)} {}^\dagger$ we find
\begin{subequations}
\begin{align}
	a = & \tilde{a}^{(2)} 
	- \lambda \frac{\hbar m^2 \omega}{4} \left(2  \tilde{a}^{(2) 3}
	- 6 \tilde{N}^{(2)} \tilde{a}^{(2) \dagger}
	+ \tilde{a}^{(2) \dagger 3} \right)
	+ \lambda^2 \frac{\hbar^2 m^4 \omega^2}{16} \left[ 2 \tilde{a}^{(2) \dagger 5} 
	+ 21 \tilde{a}^{(2) \dagger} \tilde{N}^{(2)} \tilde{a}^{(2) \dagger 2}
	- 6 \left( 23 \tilde{N}^{(2) 2} + 7\right) \tilde{a}^{(2) \dagger} \right. \nonumber \\
	& \left. 
	+ \frac{1}{2} \tilde{a}^{(2)} \left( \frac{65}{2} \tilde{N}^{(2) 3} + 27 \tilde{N}^{(2) 2} + \frac{211}{2} N^{(2)} - 9 \right)
	+ 60 \tilde{a}^{(2) 2} \tilde{N}^{(2)} \tilde{a}^{(2)} 
	+ 3 \tilde{a}^{(2) 5} \right]\\
	a^\dagger = & \tilde{a}^{(2) \dagger} 
	- \lambda \frac{\hbar m^2 \omega}{4} \left(2  \tilde{a}^{(2) \dagger 3}
	- 6 \tilde{a}^{(2)} \tilde{N}^{(2)}
	+ \tilde{a}^{(2) 3} \right)
	+ \lambda^2 \frac{\hbar^2 m^4 \omega^2}{16} \left[ 2 \tilde{a}^{(2) 5} 
	+ 21 \tilde{a}^{(2) 2} \tilde{N}^{(2)} \tilde{a}^{(2)}
	- 6  \tilde{a}^{(2)} \left( 23 \tilde{N}^{(2) 2} + 7\right) \right. \nonumber \\
	& \left. 
	+ \frac{1}{2} \left( \frac{65}{2} \tilde{N}^{(2) 3} + 27 \tilde{N}^{(2) 2} + \frac{211}{2} N^{(2)} - 9 \right) \tilde{a}^{(2) \dagger}
	+ 60 \tilde{a}^{(2) \dagger} \tilde{N}^{(2)} \tilde{a}^{(2) \dagger 2} 
	+ 3 \tilde{a}^{(2) \dagger 5} \right]~.
\end{align}
\end{subequations}
For the position operator we find
\begin{multline}
	q = \tilde{q}^{(2)} - \lambda \sqrt{\frac{\hbar^3 m^3 \omega}{32}} 3 \left[ \tilde{a}^{(2) 3} - 2  (\tilde{a}^{(2)} \tilde{N}^{(2)} + \tilde{N}^{(2)} \tilde{a}^\dagger ) + \tilde{a}^{(2) \dagger 3} \right] \\
	+ \lambda^2 \sqrt{\frac{\hbar^5 m^7 \omega^3}{512}} \left\{ 5 \tilde{a}^{(2) 5}
	+ 81 \tilde{a}^{(2) 2} \tilde{N}^{(2)} \tilde{a} 
	+ \tilde{a}^{(2)} \left(\frac{65}{4} \tilde{N}^{(2) 3} - \frac{249}{2} \tilde{N}^{(2) 2} + \frac{211}{4} \tilde{N}^{(2)} - \frac{93}{2} \right) \right. \\
	\left. 
	+ \left(\frac{65}{4} \tilde{N}^{(2) 3} - \frac{249}{2} \tilde{N}^{(2) 2} + \frac{211}{4} \tilde{N}^{(2)} - \frac{93}{2}\right) \tilde{a}^{(2) \dagger}
	+ 81 \tilde{a}^{(2) \dagger} \tilde{N}^{(2)} \tilde{a}^{(2) \dagger 2} 
	+ 5 \tilde{a}^{(2) \dagger 5}\right\}
\end{multline}
As for the norm of the perturbed eigenstate, we find up to second order
\begin{equation}
	\langle n^{(2)} | n^{(2)} \rangle = 1 + \lambda^2 \frac{\hbar^2 m^4 \omega^2}{128} \left( 65 n^4 + 130 n^3 + 487 n^2 + 422 n + 156 \right)~,
\end{equation}
while for the mean value of the position, we obtain
\begin{equation}
	\frac{\langle n^{(2)}|q|n^{(2)}\rangle}{\langle n^{(2)}|n^{(2)}\rangle} = 0~.
\end{equation}
Once again for the momentum, we find
\begin{multline}
	p = \tilde{p}^{(2)} 
	+ i \lambda \sqrt{\frac{\hbar^3 m^5 \omega^3}{32}} \left(6 \tilde{a}^{(2)} \tilde{N}^{(2)} + \tilde{a}^{(2) 3} - 6 \tilde{N}^{(2)} \tilde{a}^{(2) \dagger} - \tilde{a}^{(2) \dagger 3} \right) \\
    + i \lambda^2 \sqrt{\frac{\hbar^5 m^9 \omega^5}{512}} \left[ - \tilde{a}^{(2) 5}
    - 39 \tilde{a}^{(2) 2} \tilde{N}^{(2)} \tilde{a}^{(2)}
    - \tilde{a}^{(2)} \left(\frac{65}{4} \tilde{N}^{(2) 3} + \frac{303}{2} \tilde{N}^{(2) 2} + \frac{211}{4} \tilde{N}^{(2)} + \frac{75}{2}\right) \right. \\
    \left. + \left(\frac{65}{4} \tilde{N}^{(2) 3} + \frac{303}{2} \tilde{N}^{(2) 2} + \frac{211}{4} \tilde{N}^{(2)} + \frac{75}{2} \right) \tilde{a}^{(2) \dagger}
	+ 39 \tilde{a}^{(2) \dagger }\tilde{N}{(2)} \tilde{a}^{(2) \dagger 2}
	+ \tilde{a}^{(2) \dagger 5}\right]~,
\end{multline}
for which we have
\begin{equation}
	\frac{\langle n^{(2)}|p|n^{(2)}\rangle}{\langle n^{(2)}|n^{(2)}\rangle} = 0~.
\end{equation}

\section{Conclusion}

To summarize, in this paper we have constructed corrections to the raising and lowering, \emph{i.e.} ladder operators $a^\dagger$ and $a$ for the HO with a generic selfadjoint perturbation, polynomial in position and momentum.
It is worth emphasizing that our method can be applied to potentials depending on any power of position or momentum, \emph{not} just the second.
To our knowledge, this is the first time that such a systematic study of such generic potentials has been presented. 
Also, as remarked earlier, this is necessary and sufficient for defining coherent and squeezed states for the perturbed HO \cite{Bosso2017a}.
In this approach, one defines coherent states as eigenstates of $\tilde{a}$
\begin{equation}
	\tilde{a} |\alpha\rangle = \alpha |\alpha\rangle~,
\end{equation} similar to the standard definition \cite{Glauber1963_1}.
This implies that they follow a Poisson distribution
\begin{equation}
	|\alpha\rangle = e^{-\frac{|\alpha|^2}{2}} \sum_{n=0}^\infty \frac{\alpha^n}{\sqrt{n!}} |n\rangle~.
\end{equation}
This can be shown very easily using the definition of coherent states above and the definition of $\tilde{a}$ in \eqref{def:tilde_a}.
For the particular case studied in \cite{Bosso2017a}, motivated by GUP, it was shown that this definition also guarantees that the coherent states are minimal uncertainty states, \emph{i.e.} they saturate the Schr\"odinger--Robertson uncertainty relation for position and momentum.
Similarly, for squeezed states the standard definitions using $\sim$-operators can be applied \cite{Lu1972}.
In particular, one can define such states as those which are annihilated by the following operator
\begin{equation}
	\tilde{a}_z = \tilde{a} \cosh r - e^{i \theta} \tilde{a}^\dagger \sinh r~,
\end{equation}
where $r$ and $\theta$ are two real parameters.
In this case, although the standard statistical distribution of the number states is retained, the uncertainty relation is more complicated. 
In particular for the model in \cite{Bosso2017a}, one showed that the impossibility of 
infinite squeezing in position is consistent with the existence of a minimal length.
Given the use of such states in LIGO detectors \cite{Dwyer2013_1}, this may lead to possible observable effects of Planck-scale physics in such detectors \cite{Bosso}.

In addition to these operators doing their intended job, namely raising (lowering) the perturbed eigenstates using the standard (\emph{i.e.} unperturbed) relations, $a^\dagger |n\rangle = \sqrt{n+1}|n+1 \rangle$ (and $a |n\rangle = \sqrt{n}|n-1 \rangle$), and computing the perturbed ground state in the position or momentum representation by using the definition $a|0\rangle=0$, they also allow the straightforward computation of expectation values.
To see this, we observe that $\langle n|f(\tilde{a},\tilde{a}^\dagger)|n\rangle = \langle n^{(0)}|f(a,a^\dagger)|n^{(0)}\rangle$, as obtained directly by the definitions \eqref{def:tilde_a} and \eqref{def:tilde_a_dagger}.
In other words, the expectation value of any function of $\sim$-operators on a perturbed state is equal to the expectation value on unperturbed states of the function in which the $\sim$-operators are replaced by the standard operators.
Using this, one can easily compute expectation values with the help of the following algorithm:
\begin{enumerate}
	\item write all the operators in terms of the new ladder operators;
    \item replace the new operators by the standard ones and the perturbed states by unperturbed states;
    \item compute the expectation values in this last form.
\end{enumerate}
One expects the current formalism to be applicable to a variety of systems, which can 
be modeled by a HO with perturbations, \emph{e.g.} power law magnetic and electric perturbations of the Landau Hamiltonian \cite{landau}.
This should be generalizable 
when the number of oscillators is large, or even infinite, such as in quantum field theory 
\cite{peskin}. We hope to report on this elsewhere.  

\vspace{0.2cm}
\noindent
{\bf Acknowledgment}

\vspace{0.1cm}
\noindent
This work was supported by the Natural Sciences and Engineering Research Council of Canada and the University of Lethbridge.
P.B. acknowledges support by the Deutsche Forschungsgemeinschaft (DFG) through the grant \mbox{CRC-TR~211} ``Strong-interaction matter under extreme conditions''.

\end{document}